# Laser–assisted multiphoton ionization of a hydrogen atom by electron impact

S. Ghosh Deb , S. Roy  and C. Sinha*

Department of Theoretical Physics , Indian Association for the Cultivation of Science,

Jadavpur , Kolkata – 700 032 , India

E – mail address :chand_sin @hotmail.com

**Abstract**

The dynamics of the electron impact multiphoton ionization of a hydrogen atom in the presence of an intense laser field ($e$, $n\gamma$ $e$) has been studied theoretically, with a view to comparing ( qualitatively ) the results with the recent kinematically complete experiments of Hörr et al [ Phys. Rev. Lett. , vol. **94** , 153201 , ( 2005 ) ]  for the He target. Significant laser modifications are noted in the present doubly ( DDCS ) and the fully differential cross  sections ( TDCS ). For most of the explored kinematics ( chosen in accordance with the experiment ) , the present binary peak intensity of the laser-assisted TDCS is significantly enhanced with respect to the field free ones , in agreement with the experiment but in contradiction with the existing first order theories. Importance of the multiphoton effects are also studied.

**Introduction**

Laser-assisted ( LA ) excitation and ionization of atoms / ions are now believed to be the basic underlying mechanism for the nonsequential  double ( NSDI ) or multiple ( NSMI ) ionization processes ( in strong laser fields ) , described in the framework of recollision theory [ 1 ] . It is therefore expected that the LA elastic collision , ( e, 2e ) and simultaneous electron photon excitation ( SEPE ) experiments where the collision parameters are under full control could give detailed insight into the aforesaid processes. By virtue of the increasing progress in the availability of more powerful and tunable lasers , such processes are nowadays being observed in laboratories [ 2 – 9 ] .

Apart from the theoretical interests, the motivation for such field assisted collision processes is several fold from the practical point of view. Particularly , the LA ionization of atom / ions by charged particles play a very dominant role in many applied fields , such as , plasma confinement in fusion plasma , plasma heating , high power gas lasers , gas break down , semiconductor physics [ 1 ] , etc. Further , the LA electron - atom collision allows experimental observations of different multiphoton processes [ 10 , 11 ] at relatively moderate field intensities. It also allows to measure some electron - atom / ion scattering parameters which otherwise ( in the field free case ) would not be accessible to experiments. Despite such immense importance , the details of the dynamics of the laser assisted processes is far from being well understood and is still now a subject of great challenge [ 12 ] .

Recently Hörr et al [ 11, 13, 14 ] performed a kinematically complete experiment on the ( $n\gamma e$ , $2e$ ) of He atom using a multiparticle imaging technique. They measured for the first time the fully differential cross sections ( TDCS ) as well as the doubly differential cross sections ( DDCS ) with an exchange of up to ten photons and observed significant differences ( both in shape and magnitude ) as compared to the field free ( FF ) situations. The fully differential cross sections provide the most detailed information of an ionization process since the characteristic structures of that particular process are smoothed out by the integration steps to evaluate the lower order differentials ( doubly and singly ) and the total ionization cross sections. Thus for a proper judgement of a particular theoretical model , the study of the TDCS is highly needed.

In the experiment of Hörr et al [ 11 ] , significant differences were observed ( both in shape and magnitude ) as compared to the field free situations which could not be explained in the first order quantum calculations , e. g., the FBA or the simplified model prescribed by the Authors themselves [ 11 ]. Thus the need for a more refined theory cannot be over emphasized, as was also pointed out by Hörr et al [ 11 ] . This motivated us to attempt the present work that studies the laser- assisted multiphoton ionization ( $e$ , $n\gamma e$ ) of the H atom , theoretically the most preferred target , with the exchange of ( n ~10 ) photons and compares qualitatively the results with the experimental findings [ 11 ] for the He target. The kinematical parameters for the present study is therefore chosen in accordance with the experiment [ 11, 14 ].

Regarding the theoretical situation, a number of theoretical calculations [ 15 - 23 ] exist in the literature for the laser-assisted ( e, 2e ) problem of the H and the He atoms, prior to the experiment [ 11 ]. However, these studies refer mostly to single photon absorption or emission while the experiment involves multiphoton ( ~ 10 ) exchange. Theoretically , in a LA process , the laser usually plays the role of a third body and introduces some new degrees of freedom ( e.g. intensity, frequency and polarization ). It can thus modify the collision dynamics significantly ( as also observed in the experiment ) and hence it might be possible to control the ionization process by suitable choice of the laser parameters i.e. , to enhance or suppress the ionization cross sections as per the requirement in a particular physical process.

The present model incorporates the following distinctive features that were not taken into account in the simplified classical model [ 11 ] or in the FBA [ 15 , 24 ] .

1) The target dressing in presence of the laser field is considered both in the initial and final channels.

2) Instead of the Coulomb Volkov ( CV ) , a more refined wave function , the Modified Coulomb Volkov [ 25 ] is considered for the dressed ejected electron in the combined field of the laser as well as the coulomb field of the residual target nucleus.

3) The correlation between the projectile electron and the ejected electron is taken in to account in the final channel thereby satisfying the asymptotic three body boundary condition which is one of the essential criteria for a reliable estimate of an ionization process.

4) The projectile target nucleus ( residual ) in the final channel is considered by using the Coulomb Volkov wavefunction ( CV ) instead of the plane wave Volkov one [ 11, 15 ].

**Theory**

The present study deals with the following laser-assisted ( e, 2e ) reaction :

$$e\ (\ E_i\ ,\ \vec{k}_i\ ) + H\ (\ 1s\ ) \pm n\gamma\ (\ w,\ \vec{\varepsilon}_0\ ) \rightarrow e\ (\ E_1,\ \vec{k}_1\ ) + e^-\left(E_2\ ,\ \vec{k}_2\right) + H^+ \quad (1)$$

where $E_i$, $E_1$, $E_2$ and $\vec{k}_i$, $\vec{k}_1$, $\vec{k}_2$ are the energy and the momentum of the projectile, scattered and the ejected electron respectively, $\gamma\ (\ \omega\ ,\ \vec{\varepsilon}_o\ )$ represents the laser photon with frequency $\omega$ and field strength $\varepsilon_o$, $n$ is the number of photon exchange during the process.

The laser field is treated classically and is chosen to be single mode linearly polarized and spatially homogeneous electric field represented by $\vec{\varepsilon}\ (\ t\ ) = \vec{\varepsilon}_o\ \sin\ (\ \omega\ t + \delta\ )$ corresponding to the vector potential ( in the coulomb gauge )

$\vec{A}\ (\ t\ ) = \vec{A}_0\ \sin\ (\ \omega\ t\ +\ \delta\ )$ with $\vec{A} = c\ \vec{\varepsilon}_0\ /\omega$ and $\delta$ is the initial phase of the laser field.

Within the framework of a first order theory in the interaction potential, the prior form of the transition matrix element $T_{if}$ for the process ( 1 ) is given as :

$$T_{if} = -i \int_{-\infty}^{\infty} dt \left\langle \Psi_f^- \left| V_i \right| \psi_i \right\rangle \qquad (2)$$

In the present model the projectile – laser interaction is treated to all orders while the laser – target interaction is considered in the frame work of the first order time dependent perturbation theory. Thus the initial channel asymptotic wave function $\psi_i$ ( in Eq. 2 ) is chosen as $\psi_i = \phi_i^d \chi_{k_i}$ where $\chi_{k_i}$ denotes the plane wave Volkov solution for the laser dressed incident electron with momentum $\vec{k}_i$ and $\phi_i^d$ represents the dressed wave function of the ground state hydrogen atom [25]:

$$\chi_{k_i}(\vec{r}_1, t) = (2\pi)^{-3/2} \exp[i(\vec{k}_i \cdot \vec{r}_1 + \vec{k}_i \cdot \vec{\alpha}_0 \sin(\omega t + \delta) - E_{k_i} t)] \qquad (3a)$$

$$\phi_i^d(\vec{r}_1, t) = e^{-i\vec{a} \cdot \vec{r}_1} \frac{1}{\sqrt{\pi}} e^{-i W_0 t} e^{-\lambda \vec{r}_1}[1 - \sin(\omega t + \delta) \vec{\varepsilon}_0 \cdot \vec{r}_1 (1 + \frac{r_1}{2})] \qquad (3b)$$

The factor $e^{-i\vec{a} \cdot \vec{r}_1}$ in Eq. (3b) is introduced in order to maintain the gauge consistency between the projectile Volkov wave function and the dressed target wave function [26]. $V_i$ in Eq. (2) is the perturbation potential corresponding to the asymptotic wave function $\psi_i$ in the initial channel.. The final channel wave function $\Psi_f^-$ in Eq. (2) represents the exact solution of the following three body Schrödinger equation

$$(H - E) \Psi_f^- = 0 \qquad (4)$$

For a laser assisted ionization process, one of the major difficulties lies with the proper choice of the ejected electron wave function under the combined field of the residual target nucleus and the external laser field. In the present model the laser dressed continuum wave function of the ionized electron $\phi_{K_2}$ ( with momentum $\vec{k}_2$ ) is chosen as [19, 25]

$$\Phi^d_{k_2}(\vec{r}_2, t) = \exp(-iE_{k_2}t) \exp(-i\vec{a}\cdot\vec{r}_2) \exp[i\vec{k}_2\cdot\vec{\alpha}_0 \sin(\omega t + \delta)]$$

$$\times \left[ \psi^{(-)}_{c,k_2}(\vec{r}_2) + \frac{i}{2} \sum_n \left[ \frac{\exp[i(\omega t + \delta)]}{E_n - E_{k_2} + \omega} \right. \right.$$

$$\left. - \frac{\exp[-i(\omega t + \delta)]}{E_n - E_{k_2} - \omega} \right] M_{nk_2} \psi_n(\vec{r}_2)$$

$$+ i\vec{k}_2 \cdot \vec{\alpha}_0 \sin(\omega t + \delta) \psi^{(-)}_{c,k_2}(\vec{r}_2) \bigg] \quad (5)$$

where $\psi^{(-)}_{c,k_2}(\vec{r}_2) =$

$$(2\pi)^{-3/2} C_2 \exp(i\vec{k}_2\cdot\vec{r}_2) \,_1F_1(-i\alpha_2, 1, -i(k_2 r_1 + \vec{k}_2\cdot\vec{r}_2))$$

; $\alpha_2 = Z/k_2$ ; $E_{k_2}$ being the energy of the ejected electron and the Coulomb normalization constant $C_j = \exp(\pi\alpha_j/2)\Gamma(1+i\alpha_j)$,

$M_{n,k_2} = \left\langle \psi_n \left| \vec{\varepsilon}_0 \cdot \vec{r}_2 \right| \psi^{(-)}_{c,k_2} \right\rangle$, and $\vec{\alpha}_0 = \vec{\varepsilon}_0/\omega^2$.

It may be noted that the first term of equation (5), i.e., the zeroth order result corresponds to the so called Coulomb Volkov (CV) solution. Hereafter this will be referred as CV while the full dressing term of the ejected electron will be referred as MCV (Modified Coulomb Volkov).

The final state wave function $\Psi_f^-(\vec{r}_1, \vec{r}_2)$ in Eq. (2) satisfying the exact asymptotic three body incoming wave boundary condition is approximated as:

$$\Psi_f^-(\vec{r}_1, \vec{r}_2) =$$

$$\phi^d_{K_2}(\vec{r}_2) \chi_{k_f}(\vec{r}_1) \,_1F_1(i\alpha_{12}, 1, -i(k_{12} r_{12} - \vec{k}_{12}\cdot\vec{r}_{12})) \quad (6)$$

where $\chi_{k_f}$ corresponds to the CV wave function for the scattered electron in the final state and is given by

$$\chi_{k_f}(\vec{r}_1, t) = (2\pi)^{-3/2} \exp[i(\vec{k}_f \cdot \vec{r}_1 + \vec{k}_f \cdot \vec{\alpha}_0 \sin(\omega t + \delta) - E_{k_f} t)]$$

$$C_{1\ 1}F_1[-i\alpha_1, 1, -i(k_f r_1 + \vec{k}_f \cdot \vec{r}_1)] \qquad (7)$$

with $\vec{\alpha}_1 = 1/|\vec{k}_f|$, $\alpha_{12} = 1/|\vec{k}_{12}|$, $\vec{k}_{12} = \vec{k}_1 - \vec{k}_2$

The last term in Eq. (6) containing the confluent hyper geometric function represents the coulomb correlation between the scattered ($\vec{r}_1$) and the ejected ($\vec{r}_2$) electrons.

The expressions for the triple (TDCS) and the double (DDCS) differential cross sections accompanied by the transfer of n number of photons are

$$\frac{d^3\sigma}{d\Omega_1\, d\Omega_2\, dE_2} = \frac{k_1\, k_{2i}}{k_i} |T^n_{if}|^2 \qquad (8a)$$

$$\frac{d^5\sigma}{dE_2\, d\Omega_{1_2}} = \int \frac{d^3\sigma}{dE_2\, d\Omega_1\, d\Omega_2} d\Omega_1 \qquad (8b)$$

**Results and Discussions**:

Triple (TDCS) and Double (DDCS) differential cross sections are computed for the laser-assisted multiphoton ionization ($n\gamma\ e, 2e$) process of the hydrogen atom, where n denotes the number of photons absorbed or emitted. With a view to qualitative comparison with the only available experiment of Hörr et al [11] for the Helium target, the present kinematics are chosen in accordance with the measurement [11] and as such only the asymmetric coplanar geometry is considered. The polarization of the laser field is chosen to be parallel to the incident (electron) momentum direction. The corresponding field free (FF) results are also presented for comparison.

Figs. 1 & 2 demonstrate the angular distributions of the ejected electron (TDCS) and the difference spectra field assisted – field free (FA – FF) for the incident electron energy 1000 eV with the laser frequency of 1.17 eV and a field

strength of $5.0 \times 10^9$ V / m. Although the present study is done for the simplest H atom while the measurement refers to the experimentally preferred target, He atom, the qualitative comparison between the two is supposed to be quite meaningful. As noted from the figures, the LA binary peak of the MCV TDCS is enhanced significantly as compared to the FF one , in agreement with the experimental findings [ 11 ] but in disagreement with the FBA prediction [ 15, 24 ] which shows the reverse behaviour , e. g. , the diminution of the LA binary peak intensity . Regarding the position of the peak , the present binary peak shifts slightly towards the lower ejection angle from the momentum transfer direction $\vec{q}$, indicating the signature of higher order effects. This feature also corroborates the experimental data [ 11 ] . Needless to say, the FBA does not exhibit this feature and predicts the binary and recoil peaks along the momentum transfer directions.

Further, fig. 2 ( a ) reveals that for higher ejection energy and small momentum transfer ( e.g., for $E_2$ = 18 eV , q = 1) the present CV binary peak lies slightly below the FF peak, in corroboration with the experimental findings [ 14 ], however the MCV binary peak in this case over estimates the experiment [ 14 ] . This also establishes the fact that the present MCV model is more suitable for lower ejection energies [ 19 ] while for higher ejection energies , the CV model is better suited. The comparison of the difference spectra with the experiment in figs.1 ( e – h ) and 2 ( b ) are self explanatory .

Figures 3 demonstrate the MCV doubly differential cross sections ( DDCS ) $d^2\sigma / dq_\perp dE_2$ [ Fig. 3 ( a ) ] along with the differences of such spectra [ Figs. 3 ( b ) – ( d ) ] with the field free ones ( FA - FF ) for some selected dynamics. A clear deviation is noted between the present H atom MCV results and the experiment ( for He atom ) regarding the shape of the curve. The present MCV DDCS ( figs. 3 ( a ) ) exhibit a peak between ~ 10 – 15 eV for $q_\perp$ = 0.22 while the corresponding DDCS with the CV decreases monotonically with increasing $E_2$, the latter being in qualitative agreement with the experiment [ 11 ]. However, the difference spectra FA – FF in figs. 3 (( b ) - ( d )) more or less follows the experimental behaviour.

We now come over to figs 4 & 5 which provide some information about the individual photon exchange ( n ) distribution . Fig. 4 exhibits the distribution of n in respect of DDCS while fig. 5 plots the total inelasticity Q' [ 11 ] of the reaction $Q'/v = ( Q \pm n \hbar \omega )/ v$, $v$ being the projectile velocity and Q is the change of the total internal energy of the target. Figs. 4 & 5 may be compared qualitatively with figures 3 ( a ) & 3 ( b ) of Ref [ 11 ] .

Fig. 6 represents the difference in the DDCS spectrum ( FA – FF ) with respect to the variation of the transverse momentum transfer ( $q_\perp$ ). A fair qualitative agreement is noted with the experiment except for very low $q_\perp$ values ( ~ 0 – 0.2 a.u. ).

Finally we present ( in fig. 7 ) some additional results for the individual photon distribution ( TDCS ) at a lower incident energy 200 eV along with the high energy results ( 1000 eV ) for comparison. The importance of the former has already been emphasised by Hörr et al [ 11 ] anticipating future experiments at lower incident energies. Since for the present kinematics , the absorption ( n = 1 ) and emission ( n = -1 ) cross sections are found to be more or less similar ( both in nature and magnitude) , we present only the absorption cross sections . For lower incident energies, major modifications are noted in the LA cross sections , e.g., the FF binary peak is highly suppressed in presence of the field while an enhancement is noted in the recoil peak intensity. In other words, the ejected electron suffers a strong backward scattering in presence of the field leading to a strong recoil peak as compared to the FF case except for the no photon transfer case ( n = 0 ) where strong enhancement is noted in the LA binary peak ( not shown in the figure ). The enhancement in the present LA recoil peak ( for n $\neq$ 0 ) is in accordance with the theoretical findings of Cavaliere et al [ 15, 24 ] for single photon transfer.

In particular , for the single photon case ($n = \pm 1$) , the ratio of the binary peak intensity to the recoil peak intensity ( b / r ) changes drastically in presence of the field from $b / r \approx$ 1.17 ( FF ) to $b / r \approx$ 0.46 ( MCV ). However the CV result ( b / r $\approx$ 1.3 ) almost follows the behaviour of the FF one. On the other hand, for the two photon transfer case ($n = \pm 2$), the FF $b/r$ ratio changes from $> 1$ to $< 1$ both in the the CV and MCV models, although the change is much more significant in the latter case.

**Conclusions:**

A strong enhancement is noted in the present LA binary peak intensity of the TDCS for exchange of multi photons ($n \sim 10$) corroborating (qualitatively) the experimental findings due to Hörr et al [11].

The present MCV model that accounts for a refined wave function of the ejected electron (than the CV) gives better agreement with the experiment for lower ejection energies while for higher ejection energies the CV results are more close (qualitatively) to the experiment.

Comparison of the individual photon distributions with the multiphoton case reflects the importance of the higher values of the photon number in the LA (e, 2e) process.

The good (qualitative) agreement of the present findings with the measurement might stimulate further detailed experimental studies.

Studies on the multiphoton ionization ($e, n\gamma e$) of He atom, the experimental target, is in progress and will be reported in a future work.

**Figure Captions :**

**FIG. 1 .** The triply differential cross-section ( TDCS = $d^3\sigma / d\Omega_1 d\Omega_2 dE_2$ ) in a.u. as a function of the ejected electron angle for the ionization of hydrogen atom by electron impact in the presence of laser field . The incident projectile energy $E_i$ = 1000 eV. Laser parameters used $\varepsilon_o = 5 \times 10^9$ V / m and $\omega$ = 1.17 eV. ( Ti : Sapphire laser ) Figs. ( a – d ) the field assisted ( FA ) and field free ( FF ) TDCS .Figs ( e – h ) the corresponding difference FA – FF of the TDCS in coplanar geometry for different momentum transfer q and electron energies $E_2$.

( a ) $E_2$ = 3.7 $eV$, $q$ = 1,( b ) $E_2$ = 12 $eV$, $q$ = 1,

( c ) $E_2$ = 15 $eV$, $q$ = 1 , ( d ) $E_2$ = 15 $eV$, $q$ = 1.5 .

The figs. ( a – h ) are scaled up or down conveniently to fit the experimental data . Fig. ( a ) scaled up by 2 factor , figs. ( b & c ) scaled down by 2 factor , fig. ( f ) scaled down by 8 factor , figs. ( g & h ) scaled downed by 7 factor .

Solid curve: denotes the summed TDCS over n upto $n = \pm 3$ for MCV case, Dotted curve: denotes the same TDCS for CV case ; Dashed curve: ●denotes the FF TDCS for the ejected electron ○ . represents exp. FA and represents exp. FF [ 11 ] .

**Fig. 2** Same as fig. 1 but for $E_2 = 18$ eV and $q = 0.5$. ○ represents exp. FA and ● represents the exp. FF [11].

**Fig. 3** The doubly differential cross-sections (DDCS) in a.u. $d^2\sigma / dq_\perp dE_2$ as function of ejected electron energy ($E_2$) for different transverse momentum transfers $q_\perp$. Fig. (a) the FA DDCS considering both the full dressing of the ejected electron (MCV) and the zeroth – order dressing (CV) for $q_\perp = 0.22$. Solid curve: denotes the MCV results. Dotted curve: denotes the CV results. Figs. (b) – (d) difference field-assisted (FA) minus field free (FF) doubly differential cross-section for $q_\perp = 0.22, 0.5, 0.6$ respectively. Solid line represents the present DDCS. The experimental points [11] are given as inset in the corresponding figs.

**Fig. 4** Ratio of MCV / CV photon distributions for $q_\perp = 0.22$

**Fig. 5** Difference of the double – differential cross-section FA – FF against Q'/ v ($Q' = Q \pm n\hbar\omega$) for $q_\perp = 0.22$. Solid curve: denotes MCV results minus FF, Dotted curve: denotes CV results minus FF.

**Fig. 6** Difference of the DDCS $d^2\sigma = d^2\sigma_{FA} - d^2\sigma_{FF}$ as function of the different transverse momentum transfer ($q_\perp$) for different electron energies $E_2$ (a) $E_2 = 6$ eV, ● represents experimental data [14] (b) $E_2 = 14$ eV, experimental data [14] given as inset. Fig. (a) scaled down for comparison with the experimental data [14]

**Fig. 7** TDCS (a.u.) versus the ejection angle with $\varepsilon_0 = 5 \times 10^9$ V / m and $\omega = 1.17$ eV. The ejected electron energy is $E_2 = 5$ eV and the scattering angle is $\theta_1 = 3^0$. Figs. (a) & (b) for single photon ($n = 1$) and two photon absorption ($n = 2$) respectively for incident energy $E_i = 200$ eV. Figs. (c) & (d) represents the same for incident energy $E_i = 1000$ eV. Solid curve: MCV results, Dotted curve: CV results, Dashed curve: FF results.

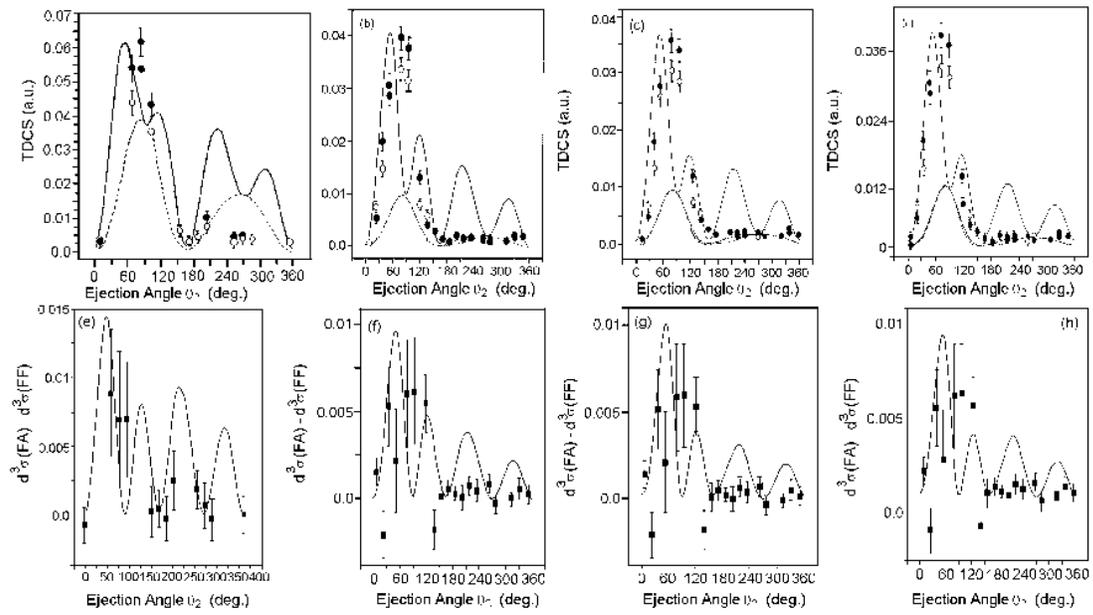

Fig . 1

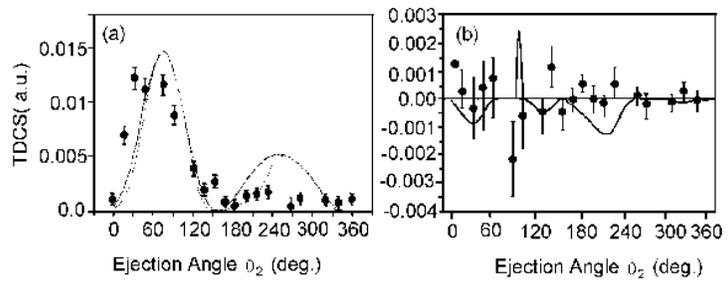

Fig . 2

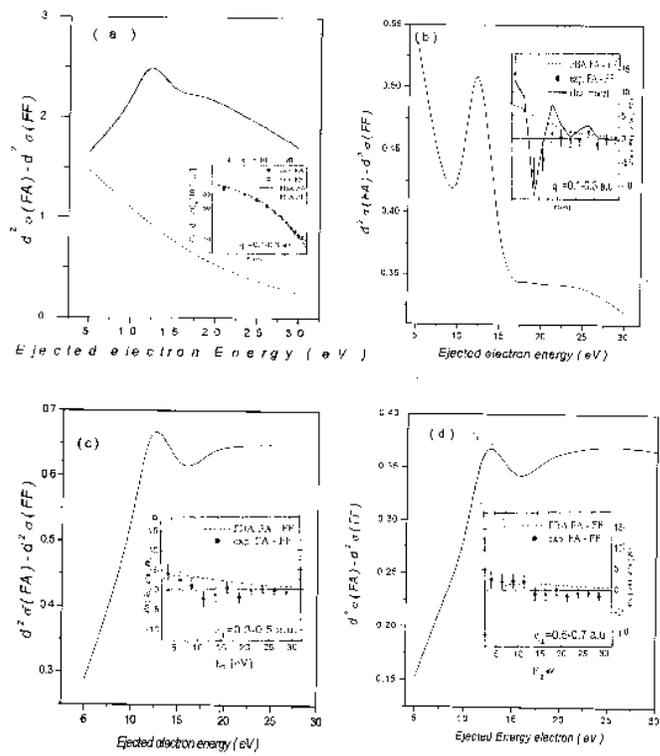

Fig. 3

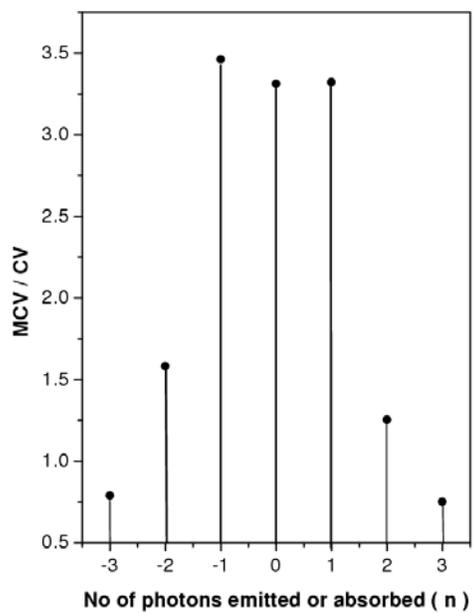

Fig. 4

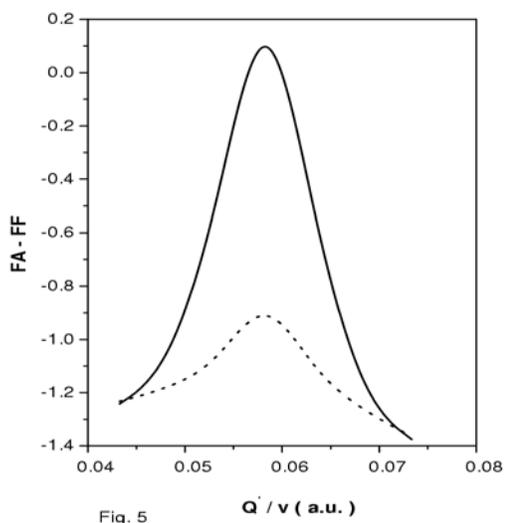

Fig. 5

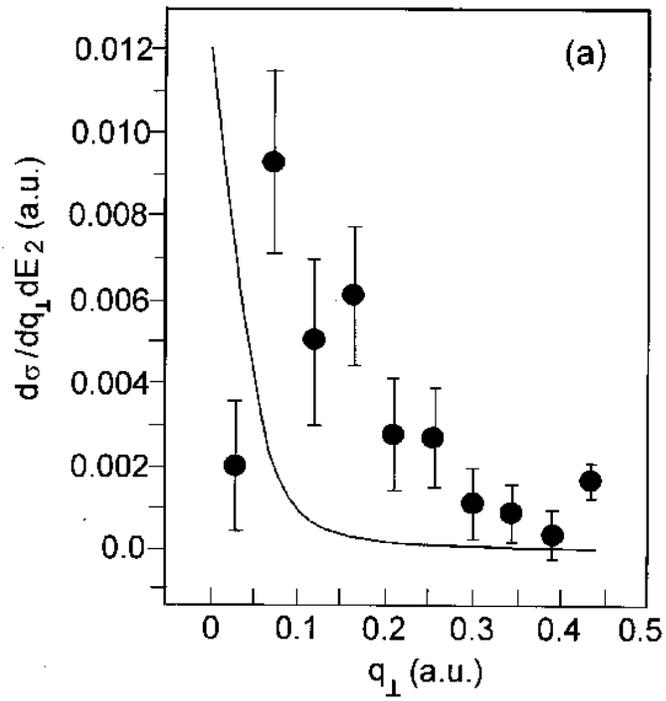

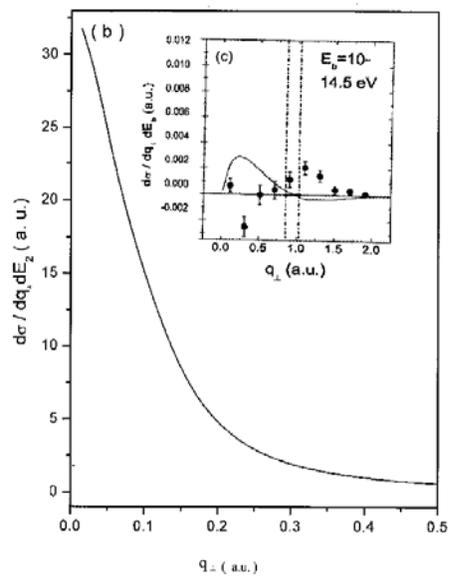

Fig. 6

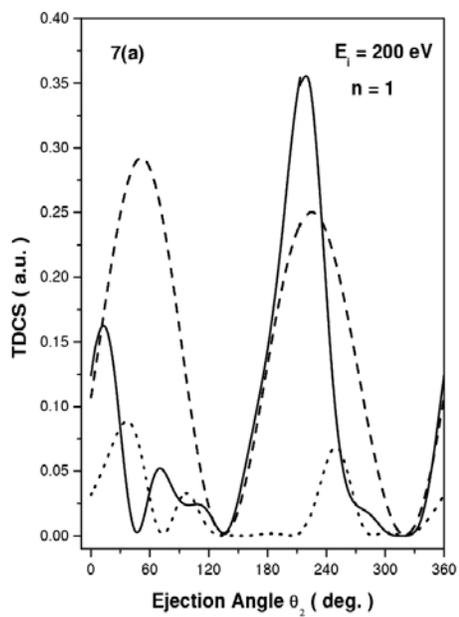

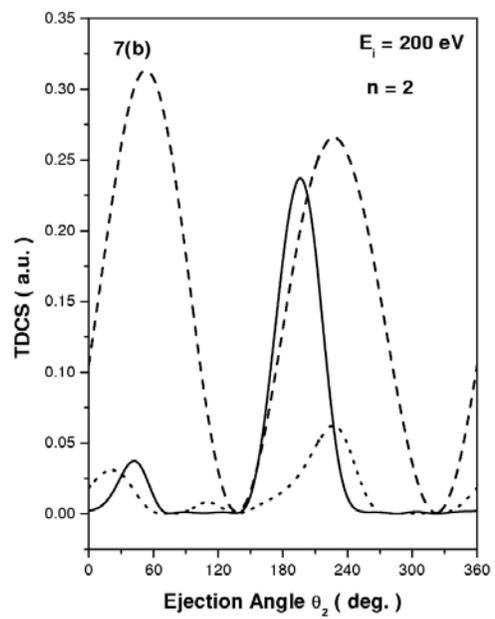

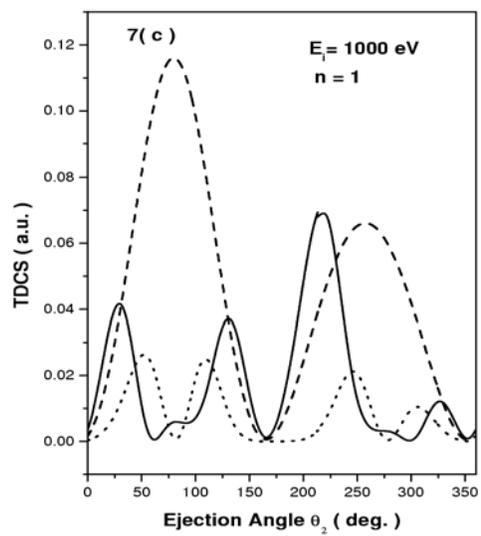

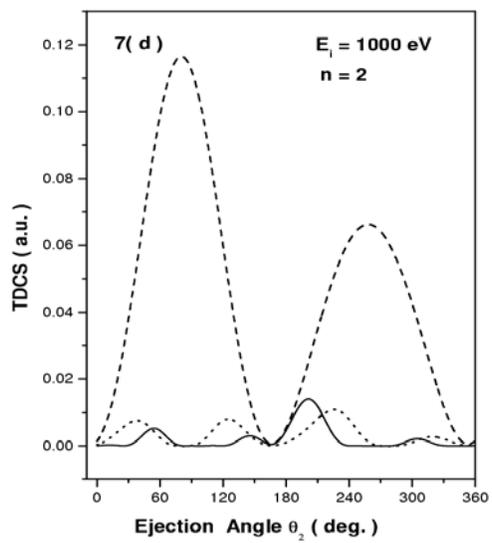